# IN VIVO FUNCTIONAL AND STRUCTURAL RETINA IMAGING USING MULTIMODAL PHOTOACOUSTIC REMOTE SENSING MICROSCOPY AND OPTICAL COHERENCE TOMOGRAPHY

*Zohreh Hosseinaee [1], Nicholas Pellegrino[1], Nima Abbasi [1], Tara Amiri[1], James A. Tummon Simmons[1], Paul Fieguth[2], Parsin Haji Reza [1]\**

1. PhotoMedicine Labs, Department of System Design Engineering, University of Waterloo, 200 University Ave W, Waterloo, ON, N2L 3G1, Canada
2. Department of System Design Engineering, University of Waterloo, 200 University Ave W, Waterloo, ON, N2L 3G1, Canada

*Email: phajireza@uwaterloo.ca

**Abstract**. We have developed a multimodal photoacoustic remote sensing (PARS) microscope combined with swept source optical coherence tomography for in vivo, non-contact retinal imaging. Building on the proven strength of multiwavelength PARS imaging, the system is applied for estimating retinal oxygen saturation in the rat retina. The capability of the technology is demonstrated by imaging both microanatomy and the microvasculature of the retina in vivo. To our knowledge this is the first time a non-contact photoacoustic imaging technique is employed for in vivo oxygen saturation measurement in the retina.

## 1. Introduction

Functional imaging techniques enable measuring alterations in biological function including metabolism, blood flow, regional chemical composition, and biochemical processes. These methods promise to dramatically improve the ability to study in situ biochemistry and disease pathology[1]. In ophthalmology, functional changes most often precede structural changes in major eye diseases. Detecting these alterations aids in understanding pathogenesis, early diagnosis, and timely management of ophthalmic disorders [2]. Retinal oxygen saturation ($SO_2$) and metabolic rate of oxygen consumption ($MRO_2$) are among the most important biomarkers characterizing the pathophysiological status of the posterior eye. Abnormal retinal $SO_2$ levels are believed to be involved in major eye diseases such as diabetic retinopathy (DR) and age related macular degeneration (AMD) [2]–[4]. Therefore, the precise measurement of retinal oxygen saturation can be critical in investigating these blinding diseases. Oxygen-sensitive electrodes and magnetic resonance imaging have been used to measure retinal $SO_2$, however these methods are usually restricted to terminal experiments and/or limited by low spatial resolution[3], [5]. Phosphorescence lifetime imaging has been also applied to map ocular oxygenation in animal models. Unfortunately, the need to inject fluorescent probes into systematic circulation makes the method inappropriate for human practice[6].

Recently, researchers are focused on optical imaging-based methods to evaluate retinal oxygenation. The advantages of optical imaging over other techniques are first the non-invasive measurement of $SO_2$ and second the ability to visualize the spatial distribution of oxygenation with high resolution in ocular environment[7]. Optical measurement of $SO_2$ is possible because the two forms of hemoglobin, oxy- and deoxyhemoglobin ($HbO_2$ and Hb), have distinct optical absorption properties. Owing to the differences in the absorption spectra of oxy- and deoxyhemoglobin, multi-wavelength imaging methods can assess the $SO_2$ in retinal vessels[8]. Currently available optical imaging modalities, such as fundus photography[9], scanning laser ophthalmoscope (SLO) [10], and optical coherence tomography (OCT) [11]–[13] are scattering based techniques and rely on the back-scattered light from the tissue to form an image. Despite the great progress in these optical imaging techniques, however, they use indirect methods to measure optical absorption. Therefore, the accuracy of these methods is affected by factors such as variation in vessel size[14], pigmentation[15], multiple light paths[16], and vessel wall thickness[8]. For example, in larger vessels the amount of detected backscattered light is much greater than in small vessels[16], hence the calculation of the optical density and $SO_2$ values can be affected in clinical trials[17], [18]. Additionally, it is shown that $SO_2$ variations, induce changes in vessel diameter[8], which may further alter the amount of backscattered light from the vessel and consequently the $SO_2$ measurements.

Photoacoustic microscopy (PAM), has the unique capability to map the direct optical absorption properties with high resolution in biological tissues[19]. The modality has the potential to overcome the limitations of current ocular imaging methods in functional studies[20]. PAM has a 100% relative sensitivity to small variations in optical absorption, which means that a given percentage of change in optical absorption coefficient yields the same percentage change in the PA signal intensity[21]. By contrast, scattering-based imaging methods has a relative sensitivity to optical absorption in blood of only ~6% at 560 nm and ~0.08 % at 800 nm[21]. It has been shown in simulation studies that by increasing the vessel diameter and melanin concentration, the relative error of $SO_2$ measurement in the scattering-based method increases. However, the $SO_2$ measurement was insensitive to these two parameters in PAM [17]. As these results suggest, PAM can be potentially a more accurate tool in quantifying retinal $SO_2$. However, the disadvantage of PAM in ophthalmic imaging arises from its need to be in contact with the ocular tissue[22]. This physical contact may increase the risk of infection and may cause patient discomfort. Additionally, it applies pressure to the eye and introduces barriers to oxygen diffusion and alter physiological and pathophysiological balance of ocular vasculature function[7]. Additionally, the use of ultrasonic signal detection scheme poses challenges in combining the technology with other imaging modalities[22].

Recently our group at PhotoMedicine Labs for the first time demonstrated non-contact photoacoustic imaging of the ocular tissue in murine eye using photoacoustic remote sensing (PARS) microscopy[23]. Later we combined the technology with OCT and applied it for *in vivo* non-contact imaging of the anterior segment in the mouse eye[7]. We showed the potential of the system for estimation oxygen saturation in the iris vasculature and visualizing melanin content in the RPE.

Our recent developments on the PARS-OCT for functional and structural imaging of anterior segment of the eye have opened a window for non-contact photoacoustic ophthalmoscopy. Building on the proven strength of PARS imaging, we further developed the technology to address the aforementioned need in retinal oximetry. In this paper we report our latest progress on in vivo retinal imaging and present the capability of the reported non-contact multimodal imaging system for visualizing structural and functional information of the rodent retina. To our knowledge, this is the first time a non-contact photoacoustic imaging technique is applied for measuring oxygen saturation in the living retina. This reported work has the potential to advance the diagnosis and treatment of major retinal disorders.

## 2. Method

2.1. System architecture

*Figure* demonstrates the experimental setup of the multimodal PARS-OCT system. Details of the system was explained in previous study published by our group for *in vivo* imaging of the anterior segment of the rodent eye[7], [23]. Briefly, the output beam of a 532-nm, ytterbium-doped laser (IPG Photonics) with 1.5 ns pulse-width is coupled to a single mode optical fiber. At the output end of the fiber multiwavelength spectral peaks are generated through stimulated Raman scattering (SRS). SRS efficiency could be affected by temperature and airflow changes, and random errors and drift in pulse energy[24]. To improve the stability of the SRS peaks, the fiber was kept in a temperature-controlled unit to isolate the airflow and the temperature[25]. The laser output was collimated into 2.1 mm diameter, then merged with the PARS probe beam centered at 830 (SLD830S-A20, Thorlabs) and the light of SS-OCT (center wavelength: 1060 nm; 100 nm spectral bandwidth, 60 kHz swept rate, Thorlabs). A two-dimensional galvanometer scanned the combined light beams (GVS012/M, Thorlabs) and relayed them to the eye through a telescopic lens (50:30 ratio) system. The pivot point of the telecentric pair is properly positioned to relay the galvanometer mirrors onto the entrance pupil of the eye, thereby minimizing vignetting. In PARS, pressure waves induced by light absorption were detected by the probe beam. The back-reflected light from the retina is directed towards the detection path. The quarter wave-plate transforms the reflected circularly polarized light back to linearly polarized light, which enables the polarized beam splitter to direct the back-reflected light towards the photodiode. The photodiode outputs are connected to a high-speed digitizer (CSE1442, Gage Applied, Lockport, IL, USA) that performs analog to digital signal conversion. In SS-OCT, backscattered light from the retina first interfered with the reference arm light, then detected by the built-in dual balanced photodetector. The OCT signal was digitized by a high-speed A/D card (ATS9351, Alazar Technologies Inc., Pointe-Claire, QC, Canada). The raw OCT data was transmitted to a host computer through a PCI-Express interface. OCT system control was implemented in MATLAB platform to automatically control all the operations including system calibration, galvo-scanning, system synchronization, real-time imaging preview and data acquisition. The lateral resolutions of both PARS and SS-OCT were around ~ 6 $\mu m$ in the retina with a slight difference caused by the different spectral ranges. The axial resolutions of PARS and OCT were ~40 $\mu m$ and ~7.3 $\mu m$, respectively.

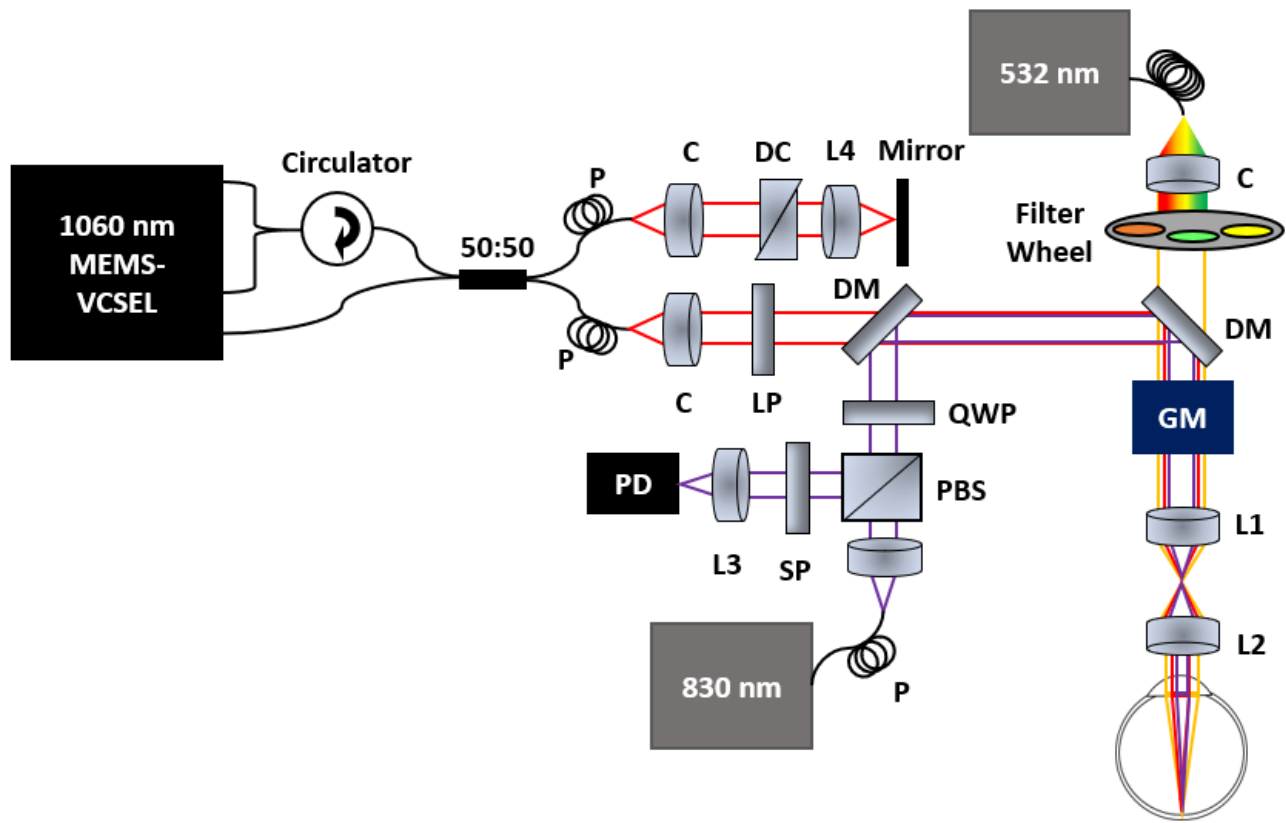

Figure 1. Simplified schematic of the multimodal PARS-OCT system. DM: Dichroic mirror, QWP: Quarter wave plate, PBS: Polarized beam splitter, LP: Long pass filter, GM: Galvanometer mirrors, L: Lens, C: Collimator, PD: Photodiode. DC: Dispersion compensation, P: Polarization controller.

2.2. Image reconstruction

All the PARS images shown in this manuscript were formed using a maximum amplitude projection (MAP) of each A-scan for each pixel of the *en-face* image. The images were produced by direct plotting from interpolated raw data using a Delaunay triangulation interpolation algorithm[26]. All images and signal processing steps were performed in the MATLAB environment.

For each OCT data set, 500 A-lines were acquired for each B-scan. For each A-line trigger, 2448 sampling points were acquired to cover the resultant spectral interferogram, providing a depth ranging distance of ~ 12 mm. To extract the OCT complex data, reference spectrum was subtracted from the interference signal to remove DC bias, then Fourier transform was performed to extract the depth-resolved OCT signal. Images were generated from the raw OCT data and numerically dispersion compensated up to the 5th order with a custom MATLAB algorithm[27]. No additional image post-processing was used for the OCT images presented in this paper. The volumetric and en-face images were generated from the 3D data sets with ImageJ[28].

2.3. Animal preparation

All the experimental procedures were carried out in conformity with the laboratory animal protocol approved by the Research Ethics Committee at the University of Waterloo and adhered to the ARVO statement for use of animals in ophthalmic and vision research. Albino rats (NU/NU, Charles River, MA, USA) were imaged to demonstrate the *in vivo* capabilities of the system. A custom-made animal holder was used to restrain the animal. The base of the animal holder was lined with a thermal pad to keep the animal body temperature between 36° and 38°C. One drop 0.5% proparacaine hydrochloride (topical anesthetic; Alcaine, Alcon, Mississauga, ON, Canada) was applied to the eye, followed by one drop of 0.5% tropicamide (pupillary dilator; Alcon). Artificial tears were used frequently (~ every 2 minutes) to keep the cornea hydrated. Vital signs, such as respiration rates, heart rates and body temperature were monitored during the experiment.

3. Results and discussion

2.4. Phantom imaging

Initial calibration of the PARS sub-system was achieved through imaging phantom eye models. First, we used a realistic human eye model to test the feasibility of PARS on the retina. The phantom is shown in *Figure 1*A, and it was modified from a commercial product (OEM-7, Ocular Instruments, Bellevue, WA). The model consists of the cornea, crystalline lens, aqueous humor, vitreous humor, and artificial retina. Strings of 7 µm carbon fibers (CF) were placed at the bottom of the model where the artificial retina is located (*Figure 1*B), and they were imaged using the PARS system. *Figure 1*C shows representative PARS image acquired from CF at the back of the human eye model and *Figure 1*D shows corresponding image acquired using PARS scattering mechanism described in our previous report[7]. The PARS excitation and PARS detection are co-aligned, so that the same fibers are in focus in both absorption and scattering contrast images (white arrow).

Then a custom rat eye model was developed to help with the alignment of PARS excitation and detection beam for in vivo trials. The eye model consisted of a single achromatic lens (63-817; Edmund Optics, Inc) with numerical aperture and curvature close to that of in the rat eye, and a custom 3D printed plastic chamber. Strings of 7 µm carbon fibers were located at a fix distance from the lens (*Figure 1*E) corresponding to the focus of the lens and was approximately equal to the path length of a rat eye [29]. Representative images acquired using absorption and scattering contrast of PARS system are shown in *Figure 1*F-G. Similar to the human eye model, the excitation and detection beams were co-aligned so that the same CF are in focus in both images (white arrows).

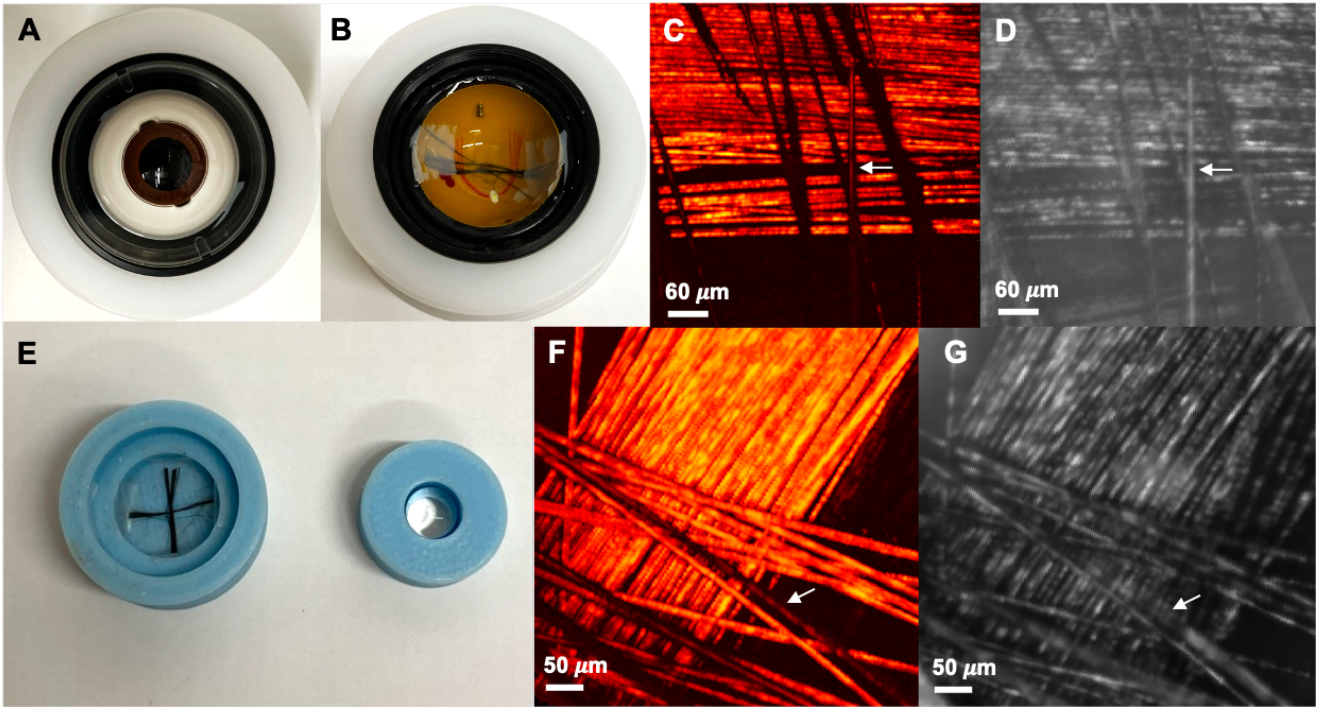

Figure 1. Imaging phantom eye models for human and rat. (A) Human eye model consisting of the cornea, crystalline lens, aqueous humor, vitreous humor, and artificial retina. (B) Strings of 7 μm carbon fibers are placed at the bottom of the eye model. (C) CF image acquired with PARS system. (D) CF image acquired using PARS scattering mechanism. (E) Custom rat eye model consisting of a single achromatic lens and a 3D printed plastic chamber and strings of carbon fibers. (F-G) PARS absorption and scattering contrast images, respectively. PARS excitation and detection beams are co-aligned so that the same CF are in focus in both images (white arrows).

2.5. $SO_2$ accuracy

The accuracy of the multiwavelength PARS sub-system in measuring oxygen saturation, was validated by performing *in vitro* phantom experiments using freshly collected bovine blood with Sodium Citrated anticoagulant solution. *Figure 2*A shows the experimental setup. The setup included a blood reservoir, tubing, oxygen tank, oxygen meter access points and a syringe pump (NE-4000, New Era Pump Systems, Inc.). Different levels of oxygen were delivered to the blood reservoir. Blood samples were drawn before and after image acquisition for CO-oximetry measurements. A clinical grade CO-oximeter (Avoximeter 4000, Instrumentation Laboratory LTD, Richmond Hill, Canada) served as a reference device for measuring blood oxygen saturation.

Images of glass capillary with flowing blood, are acquired at 532 nm and 558 nm excitation wavelength (*Figure 2*B-C). To estimate the $SO_2$, representative signal values were extracted from the two images. It is assumed that the $SO_2$ is homogeneous within the capillary and the mean values are extracted.

Compared to the per-pixel basis estimation, the mean value method offers higher accuracy by reducing the impact of random errors. Here, rather than averaging the signal intensity over the entire image, only a sub-region where the target is adequately in-focus is used. The PARS signal intensity within the region is averaged to arrive at a representative value associated with each excitation wavelength and the relative concentrations of oxy- and deoxyhemoglobin and the corresponding $SO_2$ value is calculated. The PARS measurements showed 40% and 60% oxygenation which were found to be within 6% accuracy with the oximeter results.

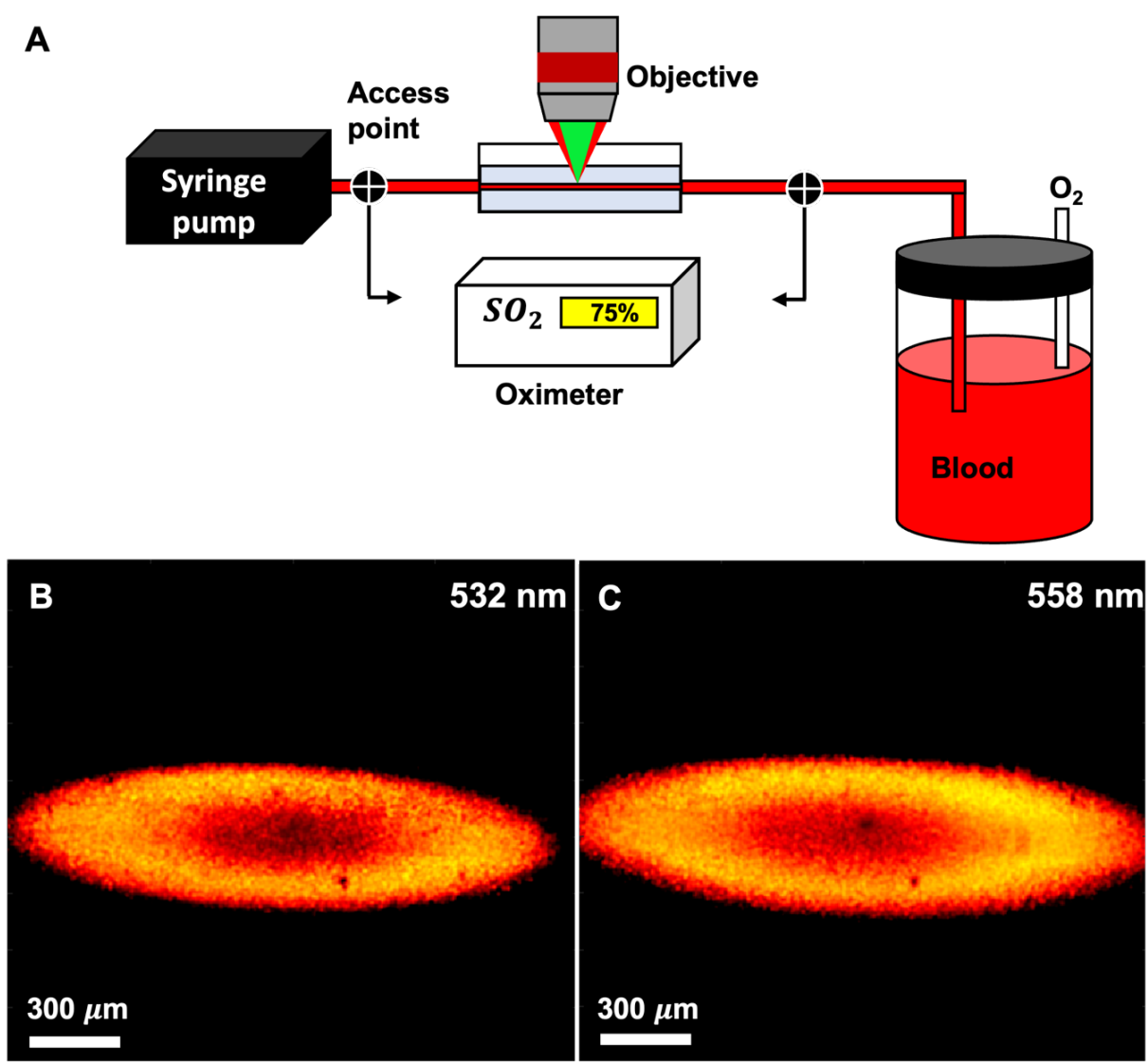

Figure 2. (A) Experimental setup of the in vitro phantom experiment using bovine blood. (B-C) Images of glass capillary with flowing blood acquired at 532 nm and 558 nm excitation wavelengths, respectively.

3.3 Retina imaging

First the capability of the SS-OCT system was tested on the rat retina, and it was used to guide the PARS imaging system. Cross-sectional and volumetric OCT images of the rat retina are demonstrated in *Figure 3*. Each data set is acquired in ~10 seconds. Cross-sectional images enable visualization of major retinal layers. Retinal nerve fiber layer (NFL) and the retinal ganglion cell layer (GCL) form the innermost (top), highly reflective band. Beneath this are the inner plexiform (IPL), inner nuclear (INL), outer plexiform (OPL), and outer nuclear layer (ONL). In principle, those layers formed of nerve fibers, i.e., IPL and OPL, show high-backscattering, whereas the nuclear layers have low-backscattering. The junction between the inner segment and outer segment of the photoreceptors (I/OS PR) could be visualized, as well as the highly reflective band which comprises the RPE (*Figure 3*B). In the OCT images central retinal artery (CRA) remnant were also visible. (*Figure 3*A). OCT fundus views are also generated by axially summing the merged OCT data set (*Figure 3*C-E). From the OCT fundus image, as shown in *Figure 3* (C), specific anatomy of the rat retina are visualized, including the optic nerve head, some large retinal vessels and optic nerve fiber bundles as indicated by the yellow arrows. *Figure 3*D show the retinal microvasculature in the deeper retinal layer (Red arrows).

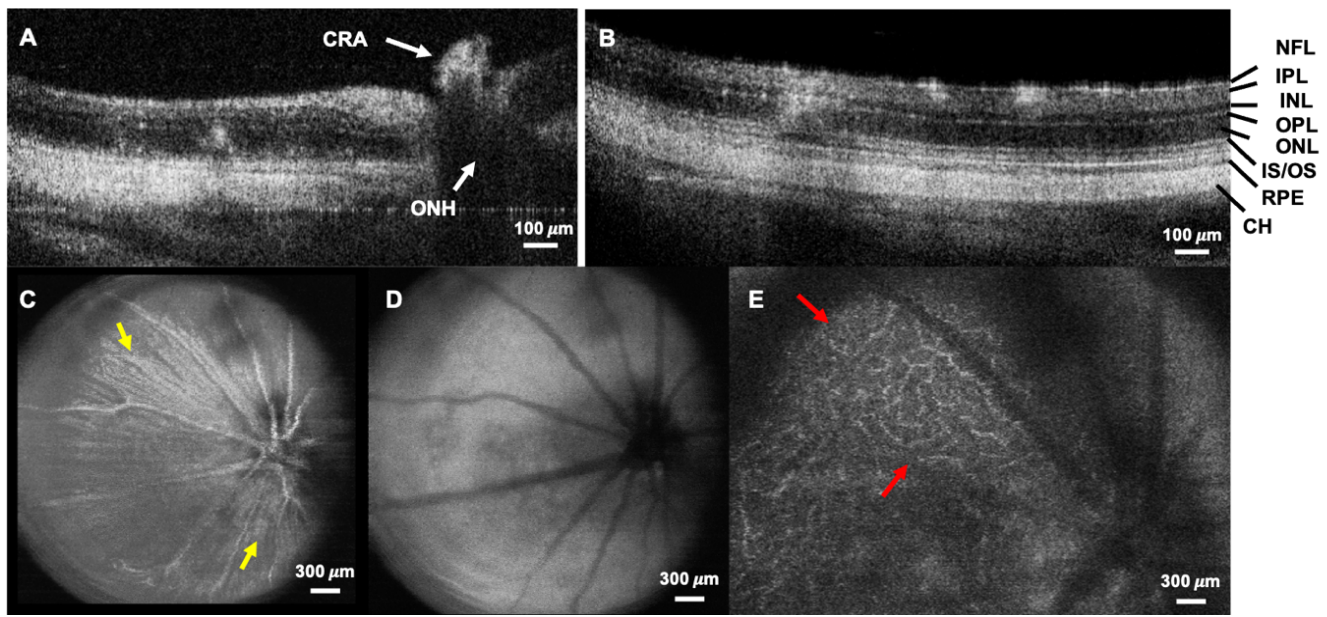

Figure 3. Volumetric and cross-sectional OCT images. (A-B) cross-sectional images acquired in vivo from rat retina showing distinct layers of the retina. NFL: nerve fiber layer, IPL: inner plexiform layer, INL: inner nuclear layer, OPL: outer plexiform layer, ONL: Outer nuclear layer, IS/OS junction of inner segment and outer segment layer. RPE retinal pigment epithelium layer, CH: choroid, CRA: central retinal artery, ONH: optic nerve head. (C-E). OCT fundus images visualizing optic nerve head, large retinal vessels, optic nerve fiber bundle (yellow arrows), deeper retinal layer microvasculature (red arrows).

These high-definition structural information of the SS-OCT subsystem can be well complemented by absorption based functional information obtained from PARS microscopy. The axial resolution of the PARS system (~ 40 $\mu$m), enables separating the signals of retinal blood vasculature from the RPE melanin given the retina thickness of ~ 200 $\mu$m[16]. As a result, PARS signals generated from major vessels can be considered to be mainly from hemoglobin. Figure 4A demonstrates fundus PARS image acquired from large vessels around ONH from a 2.6 mm × 2.6 mm area. Smaller vessels are also visible in the image; however, they are slightly distorted by motion artifacts. Figure 4B shows a zoomed-in section of one of the vessels acquired from a similar area where the smaller vasculature is more visible. In the previous report of PARS-OCT system[7], the PARS scattering contrast provided through the PARS detection beam was introduced. Figure 4C, show fundus image acquired from the ONH using scattering channel of the PARS system. The orange arrows show the vessel walls visible in the image.

Relying on the proven strength of the PARS sub-system in providing functional information, the amplitude of multiwavelength PARS signals were employed to estimate $SO_2$ in every vessel (Figure 4D). The calculation is done based on the molecular extinction coefficients of $HbO_2$ and $Hb$ at 532 nm and 558 nm optical wavelengths. The rodent's venous $SO_2$ measured with PARS was around 70% which agrees with previous reports [30]. Despite the good match between measured $SO_2$ by PARS and the preset $SO_2$ for blood experiments, the *in vivo* rodent experiments conditions are different from those in *in vitro* experiments. For example, there are eye movements during imaging, notable chromatic aberration from the eyeball, etc. Whether such differences will affect the $SO_2$ quantification or not require future validation [31], [32]. Additionally, it is reported that visual stimulus might increase the retinal vessel diameter and blood flow, as well as affect the retinal $SO_2$ due to retinal neurovascular coupling[33]. These factors may affect the accuracy of the $SO_2$ measurements, and they need further investigation in the future. To our knowledge, this is the first report of optical absorption based retinal SO$_2$ demonstrated in the fundus view.

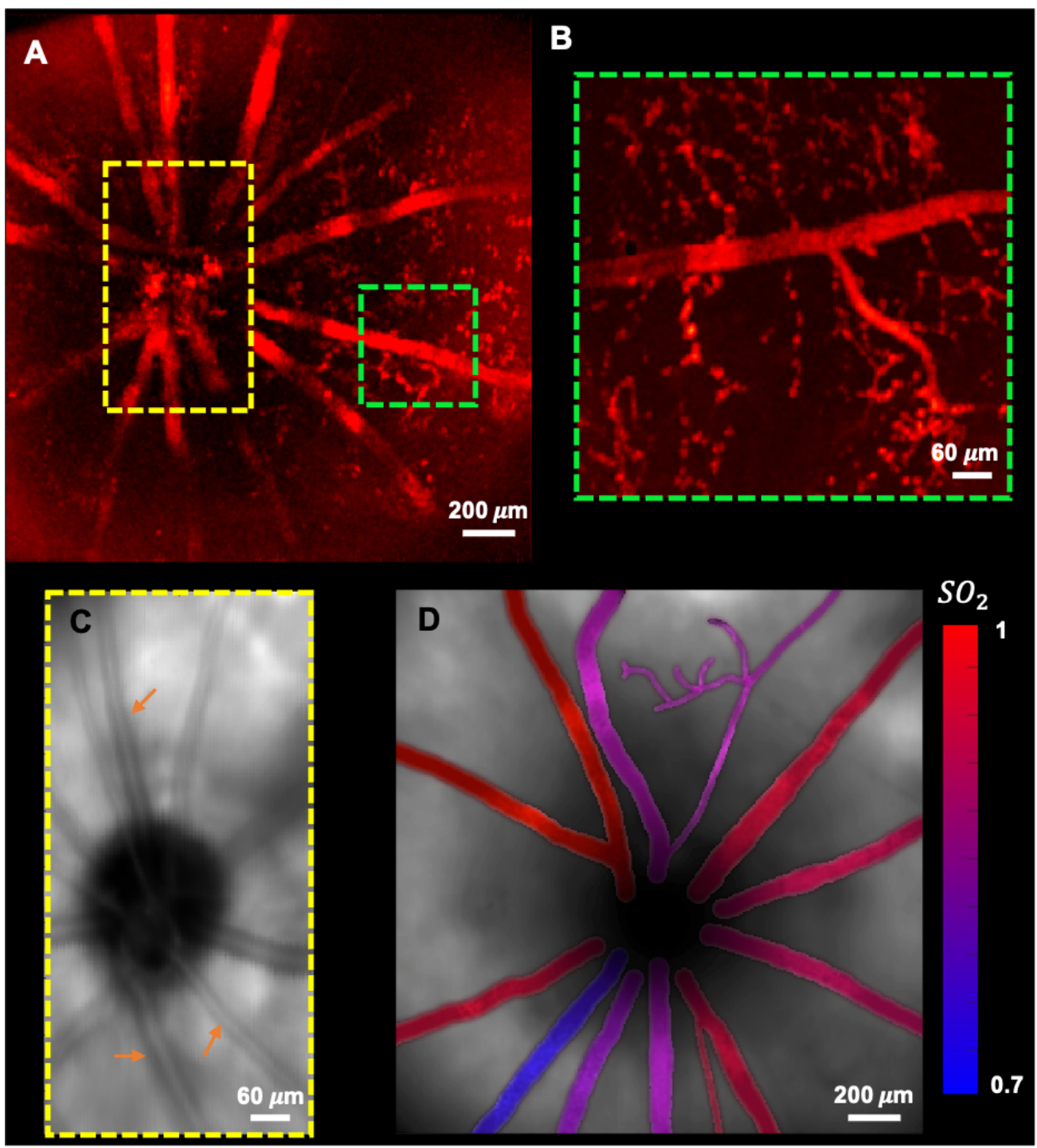

Figure 4. (A) Fundus PARS image acquired from large vessels around ONH from a 2.6 mm × 2.6 mm area. (B) zoomed-in section of one of the vessels acquired from a similar area with smaller vasculature. (C) Fundus image acquired using scattering contrast of PARS system. (D) Oxygen saturation map in the retina obtained using multiwavelength PARS imaging.

3.4. Ocular light Safety

Using ANSI standards, the ocular light safety for the developed multimodal PARS-OCT system can be calculated[34], [35]. First, the safety of PARS excitation laser is tested using the three ANSI rules: single pulse limit, average power limit, and multiple pulse limit. The pulse duration for the IPG laser used in this study is $t_1 = 1.5\ ns$. In the imaging setup, assuming that the collimated laser beam after being focused by the eye has a maximum diameter of 20 $\mu m$ on the retina, the angular subtense of the source:

$$\alpha = \frac{20\ \mu m}{17\ mm} = 1.2\ mrad < \alpha_{min} = 1.5\ mrad$$

Where 17 mm is the focal length of human eye. Therefore, the light source can be considered as a point source.

***Rule 1*** Since the pulse duration of the laser is 1.5 $ns$ only the thermal effect needs to be considered and the maximum permissible exposure (MPE) will be:

$$MPE_{SP} = 5 \times 10^{-7}\ J/cm^2$$

***Rule 2*** Since the exposure time ($\sim 10\ s$) is longer than the 0.7 $s$ and the wavelength is between 400 and 600 nm, dual limits due to both thermal and photochemical effects apply here. For the photochemical effect:

$$MPE_{ph} = 4.36 \times 10^{-2}\ J/cm^2$$

For the thermal effect:

$$MPE_{th} = 1.01 \times 10^{-2} J/cm^2$$

***Rule 3*** tests whether an exposure by a long pulse of duration $nt_1$ is safe. Within a laser spot of 20 $\mu m$, there are at most two overlapping laser pulses (n = 2).

$$MPE_{rp} = n^{-0.25} \times MPE_{sp} = 0.84 \times 5 \times 10^{-7} = 4.2 \times 10^{-7} J/cm^2$$

Rule 3 is the most conservative of the three. Considering a pupil diameter of $D = 0.7\ cm$ MPE for a single pulse would be equal to $MPE_{rP} \times \left(\frac{D}{2}\right)^2 \times \pi \approx 160\ nJ$. The acquired value is in correspondence to the values reported by other groups[36], [37]. In this manuscript the energy of a single pulse is $< 150\ nJ$ which is below the allowed pulse exposure limit.

For $\lambda$=380–1400 nm, a spectrally flat limit was recently introduced that recommends MP corneal irradiances of $25t^{-0.75}$ W/cm² for $t < 10\ s$, and 4.0 W/cm² for $t > 10s$[38]. Based on the proposed limit the OCT light power on the cornea ($\sim 1.5\ mW$) and the PARS detection power ($\sim 2 - 3\ mW$) are well within the ANSI limits.

## 4. Conclusion

Here for the first time, multimodal *in vivo* imaging of rat retina was demonstrated using multiwavelength PARS-OCT. The detailed structural information of OCT is well complemented with absorption-based functional information of PARS imaging. The proposed method can be a measure step toward non-invasive measurement of metabolic rate of oxygen consumption in retina, and it can further improve the diagnosis and treatment of important eye diseases. It was also the first time that a non-contact photoacoustic retinal imaging technique was applied for *in vivo* measuring of retinal SO$_2$. There are several aspects of the proposed system that can be further refined for future studies. The current PARS-OCT system was developed using visible excitation wavelength for targeting oxygenated and de-oxygenated hemoglobin in the blood. However, the laser safety threshold for ocular imaging is stricter in the visible spectrum than in the NIR spectral range. Additionally, the retina is sensitive to the visible light which poses challenges to eye fixation during imaging. Therefore, in future studies, the NIR spectral range can be examined as the excitation wavelength to overcome these issues.

Additionally, the scanning pattern of the SS-OCT system can be modified to enable Doppler OCT and OCT angiography and provide blood flow measurement and vasculature map, respectively. This information can be further combined with the non-contact oxygen saturation measurements of PARS to measure metabolic rate of oxygen consumption in the ocular environment in both small and large vessels. It can be further applied for evaluating the effects of visually evoked stimulus and how they could change this functional information. Furthermore, to compare the accuracy and reliability of PARS oximetry measurement in ocular tissue, it can be compared with other common methods such as visible OCT, and multiwavelength fundus photography. The system can also be used in longitudinal studies on larger animal models like rabbit and monkeys with size of eyeballs closer to that of humans. Different eye disease models (such as Glaucoma, and DR) can be developed in these animals and the efficacy of the system be evaluated in diagnosis and monitoring of them.

*Acknowledgements:*

The authors would like to thank Jean Flanagan for assistance with animal-related procedures. The authors would also like to thank Kevan Bell, Benjamin Ecclestone for their support and help. The authors acknowledge funding from New Frontiers in Research Fund – Exploration (NFRFE-2019-01012); Natural Sciences and Engineering Research Council of Canada (DGECR-2019-00143, RGPIN2019-06134); Canada Foundation for Innovation (JELF #38000); Mitacs (IT13594); Centre for Bioengineering and Biotechnology (CBB Seed fund); University of Waterloo and illumiSonics (SRA #083181).

*Additional Information*

*Conflict of interest:*

Author P. Haji Reza has financial interests in illumiSonics Inc. IllumiSonics partially supported this work. All other authors have no competing/conflict of interest.